\documentclass[
    ,final            
  ]
  {aipproc}

\layoutstyle{6x9}

\newcommand{\smpara}{{\scriptscriptstyle \parallel}}
\newcommand{\smperp}{{\scriptscriptstyle \perp}}

\begin{document}

\title{Turbulence in the Solar Corona}


\author{Steven R. Cranmer}{
address={Harvard-Smithsonian Center for Astrophysics,
60 Garden Street, Cambridge, MA 02138, USA}
}

\begin{abstract}
The solar corona has been revealed in the past decade to be
a highly dynamic nonequilibrium plasma environment.  Both the
loop-filled coronal base and the extended acceleration region
of the solar wind appear to be strongly turbulent, but direct
observational evidence for a cascade of fluctuation energy from
large to small scales is lacking.  In this paper I will
review the observations of wavelike motions in the corona over
a wide range of scales, as well as the macroscopic effects of
wave-particle interactions such as preferential ion heating.
I will also present a summary of recent theoretical modeling
efforts that seem to explain the time-steady properties of the
corona (and the fast and slow solar wind) in terms of an
anisotropic MHD cascade driven by the partial reflection of
low-frequency Alfv\'{e}n waves propagating along the superradially
expanding solar magnetic field.  Complete theoretical models
are difficult to construct, though, because many of the
proposed physical processes act on a multiplicity of spatial
scales (from centimeters to solar radii) with feedback
effects not yet well understood.  This paper is thus
a progress report on various attempts to couple these
disparate scales.
\end{abstract}

\maketitle

\section{Introduction}

Astronomers often define ``turbulence'' rather loosely, as either
a collection of motions that are unresolved either spatially or
temporally, or as dynamical oscillations that exhibit a
broad-band spectrum of frequencies and no clear dominant
frequency.
Our knowledge about turbulence in the solar corona comes
mainly from observations of this kind.
Iron-clad evidence for the existence of a turbulent cascade
in the corona awaits {\em in~situ} exploration such as
a ``Solar Probe'' could provide {\citep{Mc07}}.
Substantial progress has been made, though, on the basis of
remote-sensing observations, extrapolation inward from
existing {\em in~situ} measurements, and theoretical modeling.
This paper summarizes a cross section of these recent efforts
to better understand the role of turbulence in the corona.

\vspace*{-7.11in}
\begin{tabular}{@{\hspace*{2.10in}}p{3.40in}}
{\footnotesize To be published in proceedings of the
{\em 6th Annual IGPP International Astrophysics
Conference: Turbulence and Nonlinear Processes in
Astrophysical Plasmas,} AIP Conf.\  Ser., in press.} \\
\end{tabular}

\vspace*{6.40in}

\section{Observational Evidence}

The turbulent solar photosphere is the natural lower boundary
condition for fluctuations higher in the atmosphere
{\citep{Pt01}}.
The photosphere displays a superposition of both quasi-laminar
granulation (i.e., overturning convective cells) and smaller-scale
stochastic motions in the dark intergranular lanes.
The latter are also associated with 100~km sized concentrations of
magnetic field ($\sim$1.5 kG)---known as G-band bright points or
magnetic bright points---that are shaken transversely by the
granulation and appear to contain enough energy to give rise to
coronal Alfv\'{e}n waves {\citep{CvB05}}.

Higher in the corona, plasma fluctuations reveal themselves by
presenting variations in density, velocity, and magnetic field
strength.
These remotely measured quantities are most sensitive
to the wave modes that carry the most energy, which are generally
dominated by the longest wavelengths.
Thus, {\em direct} measurements are typically interpreted as
ideal MHD fluctuations.
(Some indirect measurements provide constraints on small-scale
kinetic modes; see below.)
The major detection techniques are listed here:
\begin{enumerate}
\item
{\bf Intensity modulations} mainly probe variations in density,
with the observed fluctuations being proportional to either $\rho$
or $\rho^2$ (integrated along the line-of-sight) depending on
the spectral band or lines used.
Intensity oscillations measured with various instruments aboard
the {\em Solar and Heliospheric Observatory} ({\em{SOHO}}) have
implied the presence of compressive MHD waves channeled along
magnetic flux tubes in coronal holes {\citep{DG98,Of00}}.
\item
{\bf Motion tracking} in sequences of images can provide information
about velocity fluctuations in the ``plane of the sky'' 
(i.e., the plane perpendicular to the line-of-sight direction).
Cross-correlation techniques have been used to obtain the bulk
solar wind acceleration of low-contrast ``blobs'' {\citep{Tp99}}.
Recently, wavelet-enhanced {\citep{SC03}} images and movies from the
EIT and LASCO instruments on {\em{SOHO}} have enabled the filtering
of fine-scale variations that were previously ``hidden'' in
diffuse larger-scale coronal features.
\item
{\bf Doppler shifts} generally allow velocities along the
line-of-sight direction to be probed.
Time-resolved sinusoidal oscillations have been seen in some
coronal structures {\citep{Wt03,Os07}}.
Most often, though, waves that reach into the solar wind are
diagnosed via ``Doppler broadening'' of spectral lines that arises
because of averaging over the oscillating redshifts and blueshifts.
In the low corona, where all plasma species are expected to be
collisionally coupled (and have identical temperatures), it is
relatively straightforward to extract the ``nonthermal'' wave
broadening from the thermal motions that also contribute to
emission line widths.
In the collisionless extended corona, though, there appears to
be at least a mild decoupling between $T_e$ and $T_p$ as well as
stronger preferential heating for heavy ions (see below).
This complicates the analysis, but realistic limits can still be
placed on transverse wave amplitudes
{\citep[for summaries of existing data, see][]{Cr04,CvB05,Es99,TM95}}.
\item
{\bf Radio sounding} probes the conditions in the corona by
measuring distortions in the net refractive index of plasma that
intervenes between a receiver and either a spacecraft beacon or
a cosmic source (e.g., a pulsar or radio galaxy).
These distortions are sensitive to density (from scintillations),
velocity (from drifting diffraction patterns), and the magnetic
field (from Faraday rotation).
The wide range of spatial scales probed by radio diagnostics
have allowed new constraints to be placed on the properties of
turbulence in the corona {\citep{Bt01,Cr02,HC05,Hroy}}, but there
is still something of a disconnect between the quantities that are
measured directly and the quantities predicted by theory.
\end{enumerate}

If the coronal fluctuations merely propagated upward on open field
lines---without interacting with the background plasma---they would
not be of much interest.
Observations of how the plasma is heated and accelerated, presumably
by {\em wave-particle interactions} in the collisionless outer
corona, are useful as an indirect means of determining which wave
modes are generated and damped.
The Ultraviolet Coronagraph Spectrometer (UVCS) on {\em{SOHO}}
measured extremely high temperatures of heavy ions, faster
ion outflow compared to protons, and strong anisotropies (in
the sense $T_{\smperp} > T_{\smpara}$) for ion
velocity distributions in the extended corona
{\cite{K97,K98,K06}}.
These properties are shared by high-speed solar wind streams
measured {\em in situ} {\citep[e.g.,][]{Ma99}}.

The UVCS observations have led to a resurgence of interest in
ion cyclotron resonance as a likely mechanism for producing this
kind of preferential ion energization, and possibly also for
heating the bulk plasma as well {\cite{Cr02,Hroy,K06}}.
Alfv\'{e}n waves that are ion cyclotron resonant in
the corona have frequencies in the $10^2$--$10^4$ Hz range,
whereas the measured and inferred frequencies are much smaller;
$10^{-5}$--$10^{-2}$ Hz.
Thus, {\em turbulent cascade} has long been considered a natural
way to produce power at high frequencies from waves initially at
lower frequencies.
It is well known, though, that both numerical simulations and
analytic descriptions of MHD turbulence (with a strong background
``guide field'' like in the corona) indicate that the cascade
from small to large wavenumber occurs most efficiently for
modes that do not increase in frequency (i.e., primarily a
fast cascade in $k_{\smperp}$ and negligible transport in
$k_{\smpara} \sim \omega / V_{\rm A}$).
In the low-$\beta$ corona, this type of quasi-two-dimensional
cascade would lead to kinetic Alfv\'{e}n waves and preferential
{\em electron} heating (in $T_{\smpara}$), which is not observed.
This issue remains a topic of active research, with several possible
outcomes depending on the behavior of the anisotropic cascade when
kinetic processes become important
{\citep{Ch05,CvB03,Hroy,Mk06,Hw06,Sc07}}.


\section{Coronal Heating and Solar Wind Acceleration}

Much of the above work dealt with determining the properties
of the ``microscopic'' fluctuations that dissipate to heat the
particles in the corona.
Complementary progress has been made in constraining the
``macroscopic'' properties of the MHD turbulence that should
eventually cascade down to the microscopic kinetic scales.
In fact, if the precise distribution of energy into various
channels (i.e., $T_{e} \neq T_{p}$ and $T_{\smpara} \neq T_{\smperp}$)
is ignored, it can be argued that the macroscopic level is
all that is needed to compute how much energy will eventually
be dissipated.
This approach has been taken in recent models of turbulent heating
of both closed loops in the low corona {\cite{Rz07}} and open
flux tubes that reach into the solar wind
{\cite{CvB05,CvB07,VV07}}.

The remainder of this paper presents results from a
self-consistent treatment of coronal heating and solar wind
acceleration that used a phenomenological description of
imbalanced MHD turbulence {\cite{CvB07}}.
The only input parameters to these models were the photospheric
lower boundary conditions for the wave spectra and the radial
dependence of the background magnetic field along the flux tube.
The models are the first self-consistent solutions that
combine: (a) chromospheric heating driven by an empirically
guided acoustic wave/shock spectrum, (b) coronal heating from
Alfv\'{e}n waves that have been partially reflected, then damped
via a turbulent cascade, and (c) solar wind acceleration
from gradients of gas pressure, acoustic wave pressure, and
Alfv\'{e}n wave pressure.

The majority of heating in these models comes from the
turbulent dissipation of partially reflected Alfv\'{e}n waves
{\citep[see also][]{Mt99,Dm02,VV07}}.
Measurements of G-band bright points in the photosphere were used
to specify the Alfv\'{e}n wave power spectrum at the lower boundary.
Non-WKB wave transport equations were then solved to determine
the degree of linear reflection due to radial gradients in the
background plasma parameters (mainly the Alfv\'{e}n speed $V_{\rm A}$).
The resulting values of the Elsasser amplitudes $Z_{\pm}$, which
denote the energy contained by upward ($Z_{-}$) and downward
($Z_{+}$) propagating waves, were then used to constrain the
energy flux in the cascade.
We used a phenomenological form for the nonlinear
transport that has evolved from studies of reduced MHD and
comparisons with numerical simulations.
The adopted volumetric heating rate (erg s$^{-1}$ cm$^{-3}$)
is given by
\begin{equation}
  Q_{A} \, = \, \rho \, \left(
  \frac{1}{1 + [ t_{\rm eddy} / t_{\rm ref} ]^{n}} \right) \,
  \frac{Z_{-}^{2} Z_{+} + Z_{+}^{2} Z_{-}}{4 L_{\smperp}}
\end{equation}
{\citep[e.g.,][]{Hs95,ZM90,Dm02,CvB07}}.
The transverse length scale $L_{\smperp}$ represents an effective
perpendicular correlation length of the turbulence, and we used
the standard assumption that $L_{\smperp}$ scales with the
cross-sectional width of the flux tube {\citep{H86}}.
The term in parentheses above is an efficiency factor that
accounts for situations when the cascade does not have time to
develop before the waves or the wind carry away the energy
{\citep{DM03}}.
The classical Kolmogorov-like cascade is thus ``quenched''
when the nonlinear eddy time scale $t_{\rm eddy}$ becomes much
longer than the macroscopic wave reflection time scale $t_{\rm ref}$ .
In most of the models we used $n=1$ based on analytic and numerical
models {\citep{Db80,Ou06}}, but we also tried $n=2$ to explore
a stronger form of this quenching.

\begin{figure}
\resizebox{.99\textwidth}{!}{\includegraphics[draft=false]{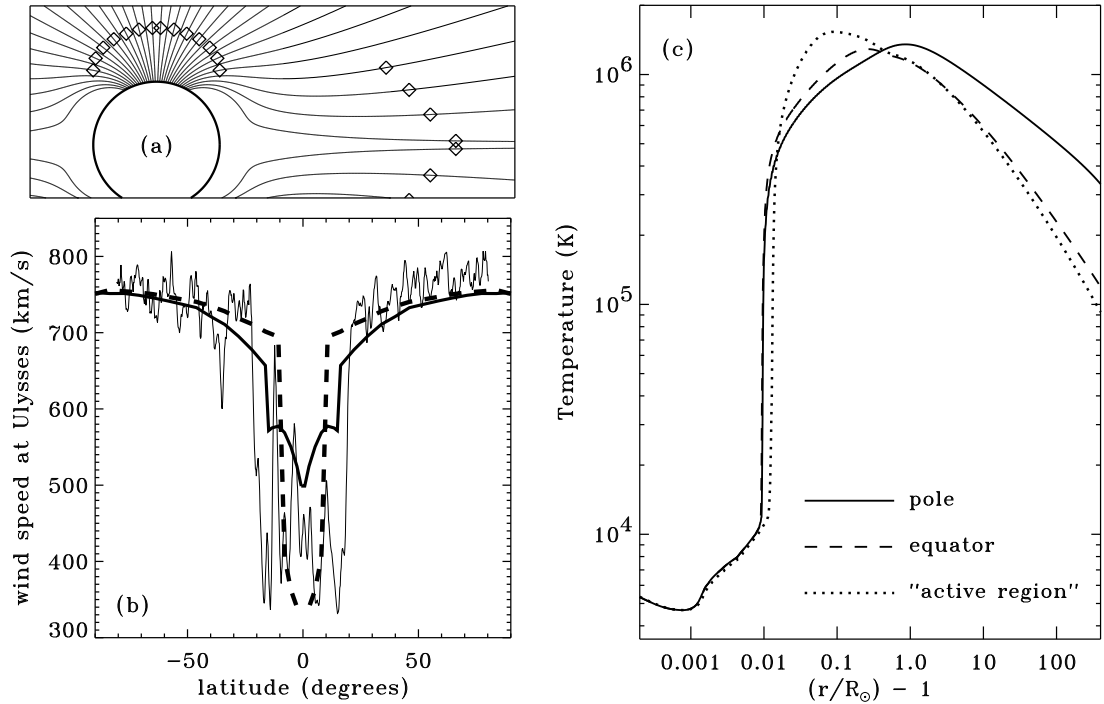}}
\caption{Summary of recent turbulence-driven coronal heating and
solar wind acceleration models.
{\bf (a)} Adopted magnetic field geometry {\citep{Bana}}, with radii
of wave-modified critical points marked by open diamonds.
{\bf (b)} Latitudinal dependence of outflow speed at $\sim$2 AU
for models with $n=1$ (thick solid curve) and $n=2$ (dashed curve),
compared with data from the {\em Ulysses} polar pass in 1994--1995
(thin solid curve) {\citep{Ge96}}.
{\bf (c)} Radial temperature dependence of polar coronal hole model
(solid curve), equatorial streamer-edge model (dashed curve), and
strong-field active region model (dotted curve).  Additional
details about these models can be found in {\citep{CvB07}}.}
\label{fig:1}
\end{figure}

Figure 1 summarizes the results of varying the magnetic field
properties while keeping the lower boundary conditions fixed
{\citep{CvB07}}.
The models included polar coronal holes, equatorial streamers
(as well as the full latitudinal variation between the two at
solar minimum) and open flux tubes rooted in active regions.
We found that a realistic variation of asymptotic solar wind
conditions can be produced by varying only the background
magnetic field geometry, as predicted by Wang and Sheeley
{\cite{WS90,WS91,WS06}}.
Specifically, the models show general agreement with some
well known empirical correlations:  i.e., a larger coronal
expansion factor gives rise to a slower and denser wind,
less intense Alfv\'{e}nic fluctuations at 1 AU, and larger
values of both the O$^{7+}$/O$^{6+}$ charge state ratio and
the FIP-sensitive Fe/O abundance ratio.
Satisfying these kinds of observational scalings are necessary
but not sufficient conditions for validating the idea that the
solar wind is driven by a combination of MHD turbulence
and non-WKB Alfv\'{e}n wave reflection.

The models shown in Figure 1 are limited by being one-dimensional,
time-independent, and one-fluid descriptions of a plasma that, in
reality, is none of those things.
Future work must involve including the divergent
temperatures and flow speeds of protons, electrons, and various
heavy ion species in the extended corona and heliosphere.
Even simple two-fluid effects ($T_{e} \neq T_{p}$) can affect
the macroscopic distribution of heat flux {\cite{HL95}},
the dynamical stability of closed-field regions like streamers
{\cite{Ev04}}, and possibly even the phenomenological form of
the MHD turbulent cascade {\cite{Gs06}}.
Also, the inclusion of compressive MHD modes may also affect many
important properties of the anisotropic cascade
{\cite[e.g.,][]{Ch05}}.

It is also likely that in certain regions of the corona,
waves and turbulence (by themselves) cannot be the whole story.
For example, it seems increasingly clear that the bright coronal
loops seen in UV and X-ray images are heated by some variety of
intermittent magnetic reconnection.
The question remains whether such a collection of bursty heating
events is powered mainly by: (a) direct stressing of the
magnetic footpoints {\cite[e.g.,][]{Gu05}}, (b) newly emerging
magnetic flux from below the photosphere {\cite{Fk03,SM03}},
(c) long-time buildup of non-potential shear {\cite{Lg04}},
or (d) some combination of the above ideas.
It is possible that concepts from turbulence theory may be
applicable to these kinds of processes as well
{\cite{vB86,Mi97}}.

\begin{theacknowledgments}
The author thanks A.\  A.\  van Ballegooijen,
R.\  J.\  Edgar, and J.\  L.\  Kohl for valuable contributions.
This work was supported by NASA under grants {NNG\-04\-GE77G,}
{NNX\-06\-AG95G,} and NAG5-11913 to the
Smithsonian Astrophysical Observatory.
\end{theacknowledgments}


\begin{thebibliography}{9}

\bibitem{Bana} M. Banaszkiewicz, W. I. Axford, and J. F. McKenzie,
  \emph{Astron.\  Astrophys.} \textbf{337}, 940 (1998).

\bibitem{Bt01} T. S. Bastian, \emph{Astrophys.\  Space Sci.}
  \textbf{277}, 107 (2001).

\bibitem{Ch05} B. D. G. Chandran, \emph{Phys.\  Rev.\  Lett.}
  \textbf{95}, 265004 (2005).

\bibitem{Cr02} S. R. Cranmer, \emph{Space Sci.\  Rev.}
  \textbf{101}, 229 (2002).

\bibitem{Cr04} S. R. Cranmer, ``Observational Aspects of Wave
  Acceleration in Open Magnetic Regions,'' in \emph{SOHO-13: Waves,
  Oscillations, and Small-scale Events in the Solar Atmosphere,}
  edited by H. Lacoste, ESA SP-547, ESA, Noordwijk, 2004, p. 353
  (astro-ph/0309676).

\bibitem{CvB03} S. R. Cranmer, and A. A. van Ballegooijen,
  \emph{Ap.\  J.} \textbf{594}, 573 (2003).

\bibitem{CvB05} S. R. Cranmer, and A. A. van Ballegooijen,
  \emph{Ap.\  J.\  Suppl.} \textbf{156}, 265 (2005).

\bibitem{CvB07} S. R. Cranmer, A. A. van Ballegooijen, and
  R. J. Edgar, \emph{Ap.\  J.\  Suppl.} \textbf{171}, in press,
  astro-ph/0703333 (2007).

\bibitem{DG98} C. E. DeForest, and J. B. Gurman, \emph{Ap.\  J.}
  \textbf{501}, L217 (1998).

\bibitem{DM03} P. Dmitruk, and W. H. Matthaeus, \emph{Ap.\  J.}
  \textbf{597}, 1097 (2003).

\bibitem{Dm02} P. Dmitruk, W. H. Matthaeus, L. J. Milano, et al.,
  \emph{Ap.\  J.} \textbf{575}, 571 (2002).

\bibitem{Db80} M. Dobrowolny, A. Mangeney, and P. Veltri,
  \emph{Phys.\  Rev.\  Lett.} \textbf{45}, 144 (1980).

\bibitem{Ev04} E. Endeve, T. E. Holzer, and E. Leer, \emph{Ap.\  J.}
  \textbf{603}, 307 (2004).

\bibitem{Es99} R. Esser, et al., \emph{Ap.\  J.} \textbf{510},
  L63 (1999).

\bibitem{Fk03} L. A. Fisk, L. A. 2003, \emph{J.\  Geophys.\  Res.}
  \textbf{108} (A4), 1157 (2003).

\bibitem{Gs06} S. Galtier, \emph{J.\  Plasma Phys.} \textbf{72},
  721 (2006).

\bibitem{Ge96} B. E. Goldstein, et al., \emph{Astron.\  Astrophys.}
  \textbf{316}, 296 (1996).

\bibitem{Gu05} B. V. Gudiksen, ``DC Heating: Is it Enough?''
  in \emph{Solar Wind 11/SOHO--16: Connecting Sun and Heliosphere,}
  edited by B. Fleck, T. Zurbuchen, \& H. Lacoste, ESA SP-592, ESA,
  Noordwijk, 2005, p. 165.

\bibitem{HL95} V. H. Hansteen, and E. Leer,
  \emph{J.\  Geophys.\  Res.} \textbf{100}, 21577 (1995).

\bibitem{HC05} J. K. Harmon, and W. A. Coles, 
  \emph{J.\  Geophys.\  Res.} \textbf{110}, A03101 (2005).

\bibitem{H86} J. V. Hollweg, \emph{J.\  Geophys.\  Res.}
  \textbf{91}, 4111 (1986).

\bibitem{Hroy} J. V. Hollweg, \emph{Phil.\  Trans.\  Roy.\  Soc.\  A}
  \textbf{364}, 505 (2006).

\bibitem{Hs95} M. Hossain, P. C. Gray, D. H. Pontius, Jr., et al.,
  \emph{Phys.\  Fluids} \textbf{7}, 2886 (1995).

\bibitem{Hw06} G. G. Howes, S. C. Cowley, W. Dorland, et al.,
  \emph{Ap.\  J.} \textbf{651}, 590 (2006).

\bibitem{K97} J. L. Kohl, et al., \emph{Solar Phys.} \textbf{175},
  613 (1997).

\bibitem{K98} J. L. Kohl, et al., \emph{Ap.\  J.} \textbf{501},
  L127 (1998).

\bibitem{K06} J. L. Kohl, G. Noci, S. R. Cranmer, and J. C.
  Raymond, \emph{Astron. Astrophys. Review} \textbf{13}, 31 (2006).

\bibitem{Lg04} D. W. Longcope, ``Quantifying Magnetic Reconnection
  and the Heat it Generates,'' in \emph{SOHO-15: Coronal Heating,}
  edited by R. W. Walsh, J. Ireland, D. Danesy, \& B. Fleck,
  ESA SP-575, ESA, Noordwijk, 2004, p. 198.

\bibitem{Mk06} S. A. Markovskii, B. J. Vasquez, C. W. Smith, and
  J. V. Hollweg, \emph{Ap.\  J.} \textbf{639}, 1177 (2006).

\bibitem{Ma99} E. Marsch, \emph{Space Sci.\  Rev.} \textbf{87},
  1 (1999).

\bibitem{Mt99} W. H. Matthaeus, G. P. Zank, S. Oughton, D. J.
  Mullan, and P. Dmitruk, \emph{Ap.\  J.} \textbf{523}, L93 (1999).

\bibitem{Mc07} D. J. McComas, et al., \emph{Rev.\  Geophys.}
  \textbf{45}, RG1004 (2007).

\bibitem{Mi97} L. J. Milano, D. O. G\'{o}mez, and P. C. H. Martens,
  \emph{Ap.\  J.} \textbf{490}, 442 (1997).

\bibitem{Of00} L. Ofman, M. Romoli, G. Poletto, G. Noci, and
 J. L. Kohl, \emph{Ap.\  J.} \textbf{529}, 592 (2000).

\bibitem{Os07} E. O'Shea, D. Banerjee, and J. G. Doyle,
  \emph{Astron.\  Astrophys.} \textbf{463}, 713 (2007).

\bibitem{Ou06} S. Oughton, P. Dmitruk, and W. H. Matthaeus,
  \emph{Phys.\  Plasmas} \textbf{13}, 042306 (2006).

\bibitem{Pt01} K. Petrovay, \emph{Space Sci.\  Rev.}
  \textbf{95}, 9 (2001).

\bibitem{Rz07} A. F. Rappazzo, M. Velli, G. Einaudi, and R. B.
  Dahlburg, 2007, \emph{Ap.\  J.} \textbf{657}, L47 (2007).

\bibitem{Sc07} A. A. Schekochihin, S. C. Cowley, W. Dorland,
  et al., \emph{Ap.\  J.\  Suppl.,} submitted,
  astro-ph/arXiv:0704.0044v1 (2007).

\bibitem{SM03} N. A. Schwadron, and D. J. McComas, \emph{Ap.\  J.}
  \textbf{599}, 1395 (2003).

\bibitem{SC03} G. Stenborg, and P. J. Cobelli,
  \emph{Astron.\  Astrophys.} \textbf{398}, 1185 (2003).

\bibitem{Tp99} S. J. Tappin, G. M. Simnett, and M. A. Lyons,
  \emph{Astron.\  Astrophys.} \textbf{350}, 302 (1999).

\bibitem{TM95} C.-Y. Tu, and E. Marsch, \emph{Space Sci.\  Rev.}
  \textbf{73}, 1 (1995).

\bibitem{vB86} A. A. van Ballegooijen, \emph{Ap.\  J.} \textbf{311},
  1001 (1986).

\bibitem{VV07} A. Verdini, and M. Velli, \emph{Ap.\  J.}
  in press, astro-ph/0702205 (2007).

\bibitem{Wt03} T. J. Wang, S. K. Solanki, W. Curdt, et al.,
  \emph{Astron.\  Astrophys.} \textbf{406}, 1105 (2003).

\bibitem{WS90} Y.-M. Wang, and N. R. Sheeley, Jr. \emph{Ap.\  J.}
  \textbf{355}, 726 (1990).

\bibitem{WS91} Y.-M. Wang, and N. R. Sheeley, Jr. \emph{Ap.\  J.}
  \textbf{372}, L45 (1991).

\bibitem{WS06} Y.-M. Wang, and N. R. Sheeley, Jr. \emph{Ap.\  J.}
  \textbf{653}, 708 (2006).

\bibitem{ZM90} Y. Zhou, and W. H. Matthaeus,
  \emph{J.\  Geophys.\  Res.} \textbf{95}, 10291 (1990).




\end{thebibliography}
\end{document}